\documentclass[dvips]{article}
\usepackage{icrctc07}

\title{Gamma-Ray Astronomy around 100 TeV with a large Muon Detector operated at Very High Altitude}
\shorttitle{Gamma-Ray Astronomy around 100 TeV}

\authors{G. Di Sciascio$^{1}$, T. Di Girolamo$^{2}$, E. Rossi$^{1}$, L. Saggese$^{2}$}
\shortauthors{G. Di Sciascio et al.}
 \afiliations{$^1$ INFN, sez. di Napoli, Italy \\$^2$ Dip. Scienze Fisiche dell'Universit\'a di Napoli and INFN, sez. di
Napoli, Italy}
 \email{giuseppe.disciascio@na.infn.it}

\abstract{Measurements at 100 TeV and above are an important goal
for the next generation of high energy $\gamma$-ray astronomy
experiments to solve the still open problem of the origin of
galactic cosmic rays. The most natural experimental solution to
detect very low radiation fluxes is provided by the Extensive Air
Shower (EAS) arrays. They benefit from a close to 90$\%$ duty
cycle and a very large field of view ($\sim$2 sr), but the
sensitivity is limited by their angular resolution and their poor
cosmic ray background discrimination. Above 10 TeV the standard
technique for rejecting the hadronic background consists in
looking for {\it "muon-poor"} showers.

In this paper we discuss the capability of a large muon detector
(A$_{\mu}$=2500 m$^2$) operated with an EAS array at very high
altitude ($>$4000 m a.s.l.) to detect $\gamma$-ray fluxes around
100 TeV. Simulation-based estimates of energy ranges and
sensitivities are presented. }

\begin{document}
\maketitle
%Begin the section.

\section{Introduction}

The recent TeV results of the HESS experiment suggest the
existence of a population of galactic $\gamma$-ray sources whose
emission extends beyond 10 TeV in the 5 to 15$\%$ of Crab flux
range (for E $>$ 1 TeV). These sources, associated with nearby
shell-type or plerionic SNRs, the most probable factories of
galactic cosmic rays, can be studied detecting gamma-rays (and
neutrinos) emission in the VHE/UHE energy domain. Therefore, a
detector capable to perform a continuous all-sky survey at a level
of about a percent of the Crab flux up to 100 TeV is needed. The
search and study of {\em "Cosmic PeVatrons"} and their surrounding
regions is one of the main scientific issues to be addressed by
the next generation of ground-based $\gamma$-ray astronomy
detectors \cite{aharonian}.

Current experiments are not able to reach 100 TeV because their
limited collection area makes the required exposures too long.
Extrapolating the galactic source spectra measured by HESS, it
appears that future experiments need to achieve at least 100
km$^2\cdot$h in order to obtain meaningful measurements at 100
TeV. The most natural experimental solution that provides such a
large exposure is given by the EAS arrays observing each source
for $\sim$1500 h/year. As an example, an experiment like ARGO-YBJ
already results in an exposure of $\sim$15 km$^2\cdot$h/year, with
an angular resolution ($\sim$0.2$^{\circ}$ at 10 TeV) near the
best value attainable by a sampling array. In addition, the EAS
arrays are the only ground-based detectors allowing simultaneous
and continuous coverage of a significant fraction of the sky
(about all that overhead). Their large field of view and high duty
cycle ($>$90 $\%$) suit to perform a $\gamma$-ray sources
population survey at VHE/UHE energies. But more important is their
unique potential that allows to have an effective monitoring of
the $\gamma$-ray activity of a large number of highly variable
sources like blazars and microquasars, as well as the possibility
of independent detection and study of GRBs \cite{aharonian}. In
addition, the recent observations of unidentified extended sources
from the Galactic plane \cite{milagro_galpl} and in the Cygnus
region \cite{milagro_cygnus} reported by the Milagro Collaboration
demonstrate the strength of EAS arrays in finding diffuse and
extended sources. Therefore, the discovery science could be a
feature of EAS arrays.

The limited sensitivity in detecting $\gamma$-ray point sources,
characteristic of EAS experiments, is mainly due to their poor
gamma/hadron separation power, limited angular resolution and high
energy threshold. Exploiting a full coverage approach at very high
altitude leads to the improvement of the angular resolution to
$\sim$0.2$^{\circ}$ and to the reduction of the energy threshold
well below the TeV region. The standard technique to perform a
gamma/hadron discrimination above 10 TeV with EAS arrays consists
in looking for {\it "muon-poor"} showers.

In this paper we discuss the capability of a large muon detector
(A$_{\mu}$=2500 m$^2$) operated with an EAS array at very high
altitude ($>$4000 m a.s.l.) to detect $\gamma$-ray fluxes up to
100 TeV. An estimation based on the simulation of energy ranges
and sensitivities is reported.

\section{$\gamma$-hadron separation}

We have simulated, via the Corsika/QGSJet code \cite{corsika},
proton- and $\gamma$-induced events with energy spectra
\cite{hegra_crab,spectra} ranging from 0.1 TeV to 1 PeV at an
observational level of 4300 m asl (606 g/cm$^2$), corresponding to
the YangBaJing Cosmic Ray Laboratory (Tibet, P.R. China). Two EAS
arrays there are located: the conventional Tibet AS$\gamma$
experiment and the full coverage detector ARGO-YBJ.
\begin{figure}
\begin{center}
\includegraphics [width=0.48\textwidth]{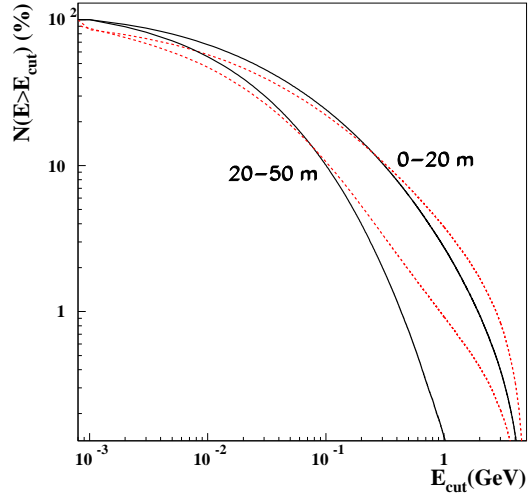}
\end{center}
\caption{Integral energy spectra of electromagnetic particles
(e$^{\pm}$, $\gamma$) produced at 4300 m asl by proton- (dotted
red lines) and $\gamma$-induced (full black lines) showers in two
different distance ranges from the core.}\label{fig:01}
\end{figure}
From the study of the shower phenomenology it results:
\begin{itemize}
\item[(1)] The muon content of proton-induced showers exceeds, as
expected, that of $\gamma$ showers by about two orders of
magnitude.
 \item[(2)] The muon component in $\gamma$-induced events has a
 flatter lateral distribution for core distances smaller than a few tens of meters.
 \item[(3)] The number of low energy e.m. particles (e$^{\pm}$,
$\gamma$) in $\gamma$ showers exceeds the corresponding number in
proton showers (see Fig. \ref{fig:01}).
 \item[(4)] Inside the core region (R$<$20 m), for particle energies above
about 400 MeV, the number of e$^{\pm}$ and $\gamma$ in proton
showers is higher that in $\gamma$ showers.
 \item[(5)] Outside the core region (R$>$20 m) the
number of e.m. particles in proton showers starts exceeding that
in $\gamma$ showers at lower energies (about 100 MeV). For E$>$1
GeV the two components differ by more than 1 order of magnitude:
high energy e.m. particles outside the core strongly indicate
proton-induced showers. This number is comparable to the number of
muons.
\end{itemize}
In addition, the high energy tail of secondary e.m. particles in
proton-induced showers is responsible for non-uniformity in the
spatial distribution of shower particles. Consequently, an
appropriate muon detector operated outside the shower core region
(R$>$30 m), evaluating all available high energy (E$>300$ MeV)
secondary components of air showers ($\mu$, e$^{\pm}$, $\gamma$),
allows a $\gamma$/hadron discrimination which goes considerably
beyond ordinary muon counting \cite{hegra_nim}. The layout for the
muon detector investigated in this preliminary study consists of 4
detectors each large 24$\times$26 m$^2$ (total area 2500 m$^2$)
simmetrically distributed on the boundary of a 150$\times$150
m$^2$ large EAS array (for example, out of the four sides of the
main building of the ARGO-YBJ experiment \cite{argo}). The
envisaged detector is constituted by 5 tracking planes shielded by
$\sim$1 m of concrete in order to absorb particles with energy
lower than $\approx$400 MeV. Therefore, in the following we will
refer to "muons" as to all the high energy particles.

%
%%%%%%%%%%%%%%%%%%%%%%%%%%%%%%%%%%%%%%%%%%%%%%%%%%%%%%%%%%%%%%%%%
\begin{figure}
\begin{center}
\includegraphics*[width=0.48\textwidth]{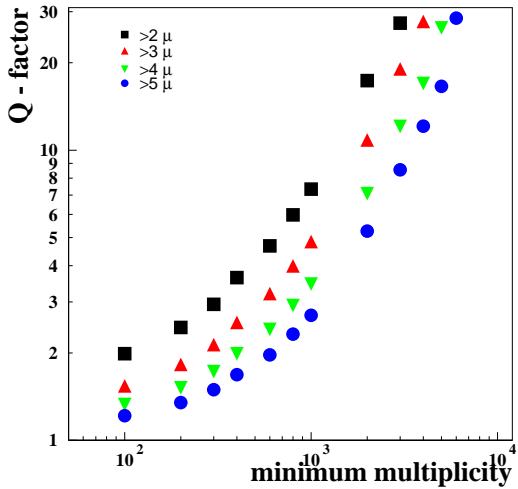}
\end{center}
 \caption{\label{fig:qfactor} Q-factor vs. minimum multiplicity
for different cuts (2, 3, 4, 5 $\mu$). Only muon detectors outside
the shower core region (R$>30$ m) have been taken into account.}
\end{figure}
%%%%%%%%%%%%%%%%%%%%%%%%%%%%%%%%%%%%%%%%%%%%%%%%%%%%%%%%%%%%%%%%%%%%
%
Compared to hadronic showers, the muon content distribution of
$\gamma$-induced events is significantly shifted toward lower
values of N$_{\mu}$ and includes a much larger fraction of events
with no muons detected. Therefore, we define as {\it muon-poor}
those events whose N$_{\mu}$ is less than a value chosen to
optimize the sensitivity to $\gamma$-ray sources. In order to
evaluate the improvement in sensitivity due to the presence of a
muon detector, we calculate the so-called {\it "Q-factor value"}
$Q_f=\epsilon_{\gamma}(r_p)/{\sqrt{1-r_p}}$, where $r_p$ is the
fraction of cosmic ray background rejected with a given N$_{\mu}$
cut and $\epsilon_{\gamma}(r_p)$ is the fraction of $\gamma$
showers which survives after this cut.

Two different analysis procedures have been performed: (1) all
muon detectors have been used in the calculation of the muon
content; (2) only muon detectors 30 m farther from the
reconstructed shower core position have been taken into account.
In Fig.\ref{fig:qfactor} the Q-factor values as a function of
multiplicity for different N$_{\mu}$ cuts are shown. Only
$\mu$-detectors outside the shower core region (R$>30$ m) have
been taken into account to calculate the muon content. The
multiplicity is the number of charged particles N$_{ch}$ sampled
by a 100$\times$100 m$^2$ full coverage detector whose 4 sides are
surrounded by the 4 $\mu$-detectors.

\section{Sensitivity to $\gamma$-ray sources}

\begin{figure*}
\vspace{-0.5cm}
\begin{minipage}[t]{.48\linewidth}
\includegraphics*[width=0.9\textwidth,angle=0,clip]{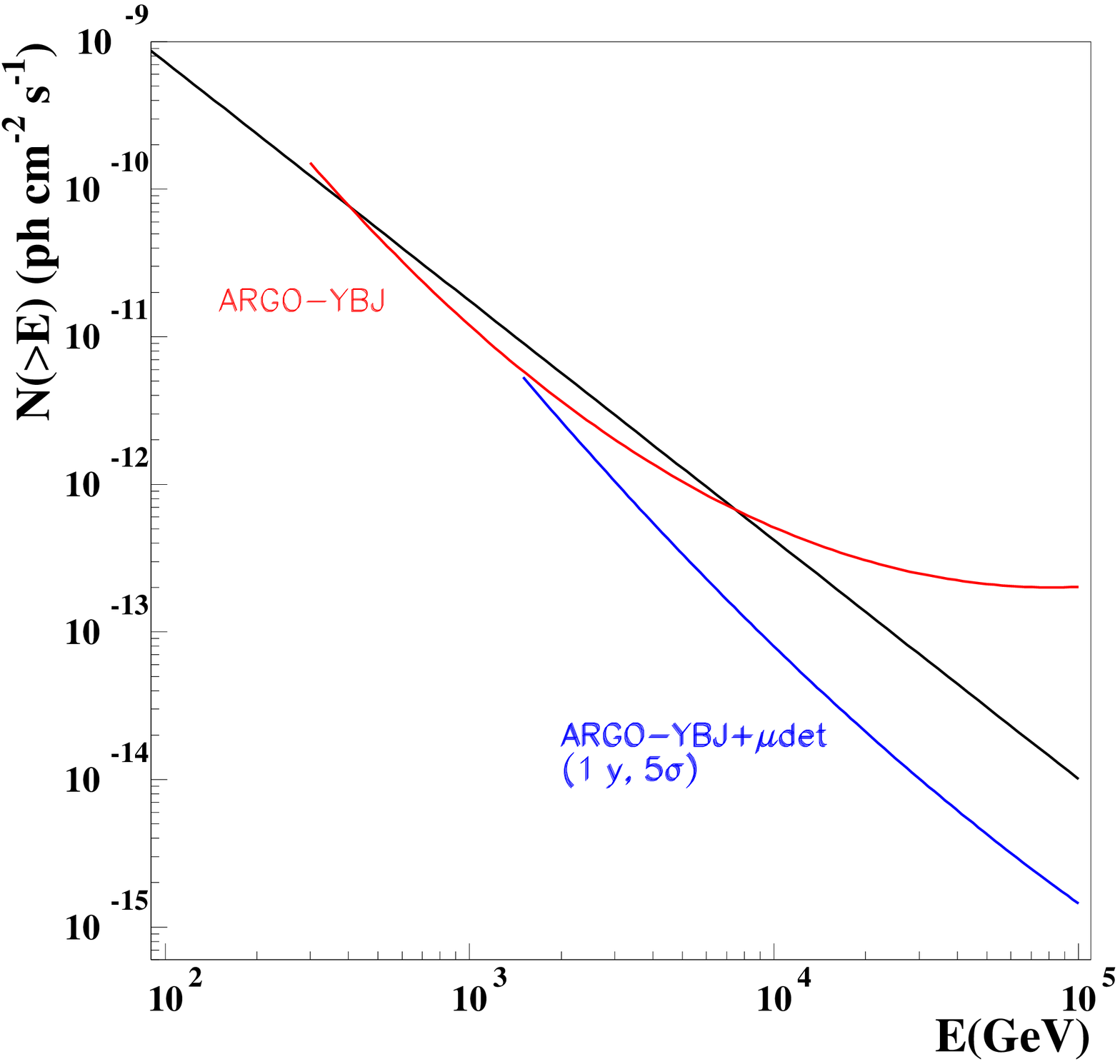}
 \caption{\label{fig:senscfr} Minimum Detectable Flux (1 year,
5$\sigma$) compared with the Crab spectrum (black line). The red
curve shows the ARGO-YBJ sensitivity, the blue one the MDF due to
a 2500 m$^2$ $\mu$-detector.}
\end{minipage}\hfill
\begin{minipage}[t]{.48\linewidth}
\includegraphics*[width=0.9\textwidth,angle=0,clip]{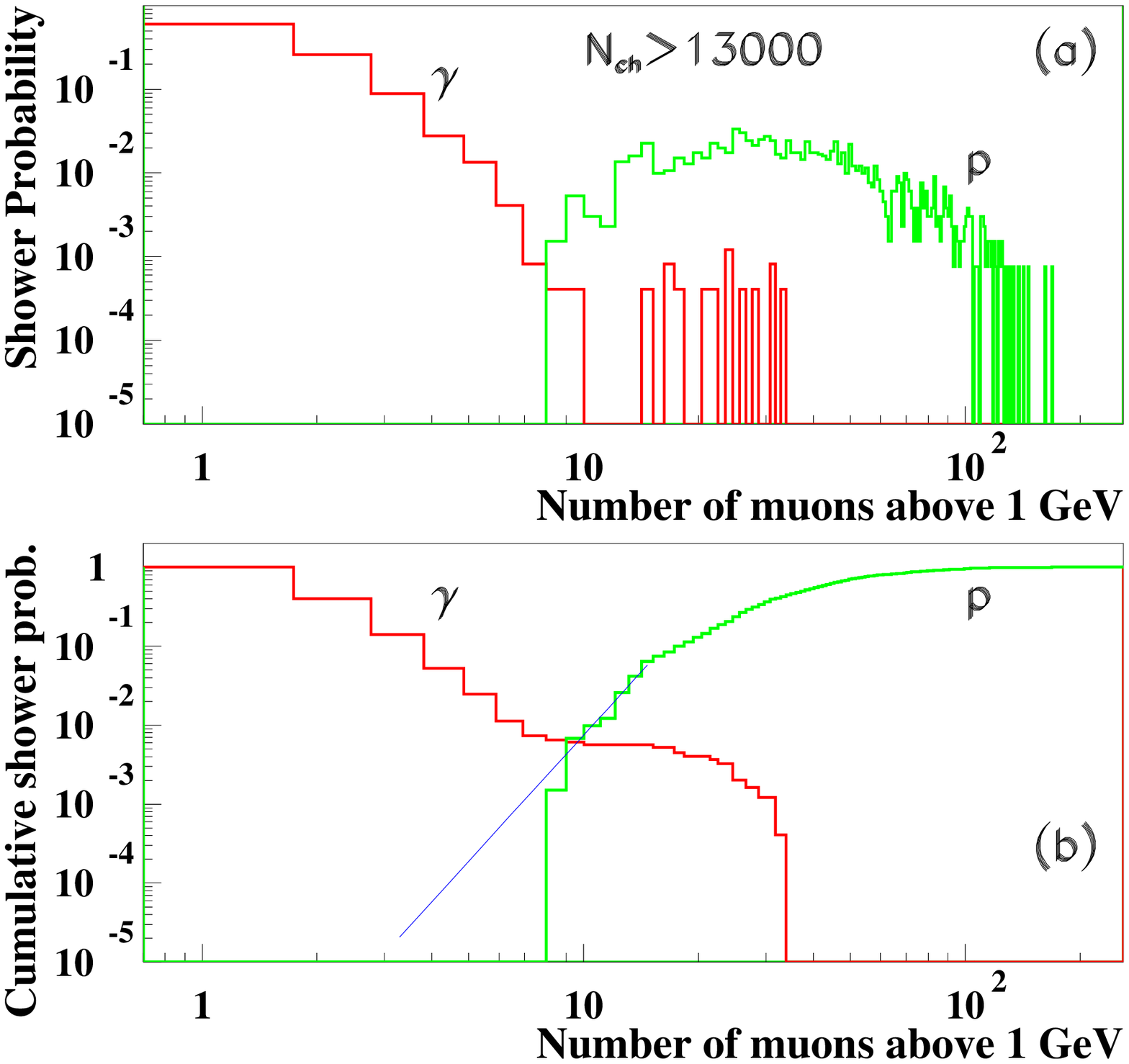}
\caption{\label{fig:rejection} (a) Probability that a photon or
proton shower generates a given number of muons. (b) Inclusive
probability for the histogram shown in (a).}
 \end{minipage}\hfill
\end{figure*}

In order to evaluate the Minimum Detectable Flux (MDF) improvement
due to the calculated Q-factors, we considered as a reference
sensitivity the ARGO-YBJ MDF calculated for internal events, at 5
standard deviations in one year of observation for a Crab-like
source \cite{argo} (red curve of Fig. \ref{fig:senscfr}). The
results of these calculations with procedure (2) are shown in
Fig.\ref{fig:senscfr} by the blue curve. For a conservative
estimation of the MDF we exclude cuts at very low (0, 1 and 2)
muon content, where there is a potential background from
mismatched events, and we consider that only the showers with $<$3
muons on the detectors are due to a $\gamma$-ray. At 30 (80) TeV
the improvement in the MDF, due to the rejection of showers with
$\geq$ 3 detected muons, is a factor 25 (100), because at
sufficiently high energy the $\gamma$-ray measurement is
background-free.

%
%%%%%%%%%%%%%%%%%%%%%%%%%%%%%%%%%%%%%%%%%%%%%%%%%%%%%%%%%%%%%%%%%
%%%%%%%%%%%%%%%%%%%%%%%%%%%%%%%%%%%%%%%%%%%%%%%%%%%%%%%%%%%%%%%%%%%%
%
%

\subsection{The rejection power for diffuse $\gamma$-rays}

The detection of an isotropic diffuse flux of UHE photons depends
not only on the size and the sophistication of the detectors but
also on the properties of the $\gamma$-ray showers. In fact, if
these events have the same muon content as hadronic showers the
detection of any diffuse $\gamma$-ray fluxes would be impossible.
In order to evaluate the rejection power for proton-induced
showers we have studied how frequently hadronic showers fluctuate
in such a way to have a low muon content indistinguishable from
$\gamma$-induced events. In Fig.\ref{fig:rejection} we show the
probability that a photon or proton shower generates a given
number of muons on the simulated detector, assuming for both
primaries the same spectral index -2.7. We note that the
fluctuations about the mean values do not follow the Poisson
statistics. The inclusive probability distributions are reported
in the lower panel of the figure. For proton (photon) showers the
histograms bars in the inclusive distributions represent the
probability that a shower contains less (more) muons than the
upper (lower) limit of the bin. The plot refers to showers with
$N_{ch}\geq$13000 (photon median energy $\sim$150 TeV). We have
conservatively fitted the proton distribution overestimating the
background in order to predict their level at very low muon
content \cite{gaisser}. In Table 1 we list the implications of
$\gamma$-ray detection for showers with $N_{ch}\geq$13000. In this
energy region a significant fraction of the cosmic rays could be
heavy nuclei, yielding larger muon numbers compared to proton
showers alone. Therefore, we are studying here a worst-case
scenario, as protons are most likely to fake $\gamma$-rays. From
MC calculations we conclude that diffuse $\gamma$-rays of $\sim$
150 TeV energy can be observed down to a level $\sim$10$^{-5}$ of
the cosmic ray background if the systematic uncertainties of the
detector are understood at the same level.
%
%%%%%%%%%%%%%%%%%%%%%%%%%%%%%%%%%%%%%%%%%%%%%%%%%%%%%%%%%%%%%%%%%%%%%%
\begin{table}
\begin{tabular}{lccc}
\hline
 $N_{\mu}$  & $<$3 & $<$4 & $<$5 \\
\hline
Fraction of $\gamma$-ray &           &                  &                  \\
retained                 & 85$\%$    & 94$\%$           & 97$\%$          \\
                         &           &                  &  \\
Background level         & $10^{-5}$ & 6$\cdot$10$^{-5}$ & 2$\cdot$10$^{-4}$ \\
 \hline
\end{tabular}
\caption{}
\end{table}
%%%%%%%%%%%%%%%%%%%%%%%%%%%%%%%%%%%%%%%%%%%%%%%%%%%%%%%%%%%%%%%%%%%%%%%%
%

\subsection{Conclusions}

Recent discoveries in the TeV $\gamma$-ray astronomy provide
strong motivations for extending the measurements to energies
$>$100 TeV where $\gamma$-ray emission models can be better
discriminated. EAS arrays naturally provide large exposures
allowing simultaneous and continuous coverage of $\sim$2 sr
without facing serious technological challenges. A suitable large
muon detector operated with an EAS array at very high altitude
permits a $\gamma$ measurement essentially background-free for E
$>$50 TeV.

\end{document}